**Литвинова Світлана Григорівна**
доктор педагогічних наук, с.н.с, завідувач відділу технологій відкритого навчального середовища
Інститут інформаційних технологій і засобів навчання НАПН України, м. Київ, Україна
ORCID ID 0000-0002-5450-6635
*s.h.lytvynova@gmail.com*

# СИСТЕМА КОМП'ЮТЕРНОГО МОДЕЛЮВАННЯ ОБ'ЄКТІВ І ПРОЦЕСІВ ТА ОСОБЛИВОСТІ ЇЇ ВИКОРИСТАННЯ В НАВЧАЛЬНОМУ ПРОЦЕСІ ЗАКЛАДІВ ЗАГАЛЬНОЇ СЕРЕДНЬОЇ ОСВІТИ

**Анотація.** У статті проаналізовано історичний аспект формування комп'ютерного моделювання як одного з перспективних напрямків розвитку освітнього процесу. Обґрунтовано поняття «система комп'ютерного моделювання», концептуальну модель системи комп'ютерного моделювання (СКМод), визначено її складові (математична, анімаційна, графічна, стратегічна), функції, принципи та цілі використання. Описано особливості організації роботи з учнями на засадах використання СКМод, індивідуальної й групової роботи, формування предметних компетентностей; розглянуто аспект мотивації учнів до навчання. Встановлено, що заклади освіти можуть використовувати СКМод на різних рівнях і етапах навчання. Визначено, що використання СКМод в закладах загальної середньої освіти збільшить можливості вчителів щодо вдосконалення організації процесу навчання і сприятиме його індивідуалізації, з метою задоволення темпів, освітніх інтересів і можливостей кожного конкретного учня. Обґрунтовано, що застосування СКМод при вивченні природничо-математичних предметів сприяє формуванню предметних компетентностей, розвиває навички аналізу і прийняття рішень, підвищує рівень цифрової комунікації, розвиває пильність, підвищує рівень знань та збільшує тривалість уваги учнів. Подальшого дослідження потребує обґрунтування процесу формування компетентностей учнів з природничо-математичних предметів та проектування пізнавальних завдань з використанням СКМод.

## 1. ВСТУП

**Постановка проблеми.** Нині в суспільстві занепокоєння викликають дві проблеми. По-перше, відсутність висококваліфікованих науковців сьогодні може стримувати науково-технічний прогрес в майбутньому. По-друге, сьогоднішнім студентам у майбутньому розуміння науки та технологій буде необхідним для прийняття обґрунтованих рішень щодо найважливіших соціально-наукових питань, починаючи від глобального потепління до особистих (медичних, освітніх, соціальних). Однак низький рівень підготовки сучасних учнів з питань природничо-математичної освіти, що є основою розвитку новітніх технологій у таких сферах як медицина, хімічна, харчова, аграрна промисловості, може поглибити ці проблеми.

Новий напрям розвитку технологій Cyber-Human Systems базується на повсюдному використанні людиною комп'ютера і конкретних методологічних підходах. Формування такої технології більшою мірою базується на комп'ютерному моделюванні різних процесів, необхідних для життєдіяльності людини, визначенні їх переваг і недоліків, здійсненні тематичного і наскрізного аналізу результатів і даних, а також візуалізації, що є одним з важливих компонентів цього процесу. Використання таких технологій вимагає відповідної підготовки як студентів, так і школярів. Використання в освітньому процесі комп'ютерного моделювання має допомогти учням закладів загальної середньої освіти глибше зрозуміти природні процеси та явища, а вчителям – активізувати їх навчальну діяльність в напрямках виконання проектних, дослідницьких, творчих і проблемних завдань. Отже у XXI ст. освітній простір набуває нових рис з посиленням ролі синтетичного навчального середовища [1].

**Аналіз останніх досліджень і публікацій**. Формування системи відкритої освіти, на засадах використання ІКТ, комп'ютерно і мобільно орієнтованих середовищ навчання та електронних освітніх ресурсів, зокрема ігрових, розкрито в працях В.Ю. Бикова [2].

Питання комп'ютерного моделювання та застосування мережі Інтернет для дослідження природних явищ обґрунтовано вченими Ю.О. Жук, О.М. Соколюк, Н.П. Дементієвською, О. В. Слободяник. Дослідники розкрили основні принципи і підходи до використання Інтернет-технологій в шкільному експерименті під час вивчення курсу фізики і констатували, що мають бути визначені конкретні навчальні цілі для застосування комп'ютерного моделювання. Такі інноваційні підходи в організації навчання спонукатимуть учнів до пошуку причинно-наслідкових зв'язків та значущих висновків. Учні зможуть залучити попередні знання, досвід та розуміння до здобуття та побудови нових знань, співставлення отриманих знань з реаліями оточуючого світу. У процесі використання комп'ютерного моделювання учитель має надавати тільки мінімальні настанови, щодо їх використання і координувати діяльність, що потребує і передбачає співпрацю учнів [3].

Дослідники з Університету імені М.П. Драгоманова (О. В. Матвійчук, В. П. Сергієнко, С. О. Подласов) встановили, що засвоєння знань учнями відбувається більш ефективно в процесі діяльності. Такою діяльністю може бути розробка комп'ютерних моделей фізичних явищ. Створення комп'ютерної моделі, перш за все, вимагає від учня глибшого розуміння сутності процесів, що відбуваються, та їх математичного описання. При цьому процес побудови комп'ютерної моделі можна організувати з поступовим її ускладненням і наближенням до реальності, що відповідає дидактичному принципу "від простого до складного" [4].

Громко Г.Ю. обґрунтовує, що для створення комп'ютерних моделей і проведення віртуального фізичного експерименту необхідні програмні засоби, наприклад Scratch. Формалізованість комп'ютерної моделі дозволяє виявити основні чинники, що визначають властивості досліджуваних об'єктів, вивчати реакцію деякої фізичної системи на зміни її параметрів. Зокрема імітаційні моделі відтворюють алгоритм функціонування досліджуваної системи шляхом послідовного виконання великої кількості елементарних операцій [5].

Хазіна С. А. додає, що педагогічно доцільне і виважене впровадження в навчальний процес комп'ютерного моделювання дає змогу забезпечити розвиток інтелектуальних умінь, глибоке розуміння процесів, що моделюються, формувати дослідницькі вміння, поглиблювати знання і вміння з інформатичних, фізичних та математичних дисциплін, удосконалювати навички роботи в різних програмних середовищах [6].

Л. В. Резниченко, зазначає, що досвід використання засобів комп'ютерного моделювання у навчанні майбутніх учителів природничих дисциплін свідчить про появу нових можливостей у процесі пізнання "хімічного світу", які не досягаються іншими традиційними засобами, а використання програми AutoDock Vina і етапність в процесі навчання дозволяє студентам чітко усвідомлювати послідовність виконання дій і наочного зображення наслідків цих дій [7].

Т. М. Деркач звертає увагу на використання середовища програмування NetLogo у навчанні хімії. Розроблені в NetLogo комп'ютерні моделі можна застосувати для організації самостійної роботи в рамках вивчення курсу неорганічної хімії [8].

У звіті групи зарубіжних науковців представлено досвід і результати використання ігрових комп’ютерних моделей у навчанні природничих предметів. Вони зазначають, що використання комп’ютерного моделювання в процесі навчання учнів середньої школи дає значний позитивний результат, що підтверджується рівнем їх компетентностей [9].

Зазначимо, що нині ми маємо результати наукових досліджень в яких розкрито позитивний досвід використання комп’ютерного моделювання за такими напрямками: комп’ютерне моделювання у підготовці майбутніх інженерів-педагогів (Р.М. Горбатюк); комп’ютерне моделювання анімаційних наочностей засобами графічного середовища програми Maxima (Н.О. Бугаєць); ігрове моделювання як засіб підвищення активності навчання (Є.В. Прокопенко); комп’ютерне засвоєння базових предметів методом імітаційного моделювання (Р.М. Павленко), моделювання як ефективний метод посилення міждисциплінарних зв’язків (М. О. Мястковська); використання імітаційного моделювання в освітньому процесі (Т.О. Фадєєва), активізація дослідницької діяльності учнів на засадах використання систем комп’ютерної математики широко представлені в працях О.О. Гриб’юк.

Однак проблема використання комп’ютерного моделювання для вивчення природничо-математичних предметів в закладах загальної середньої освіти розкрито не повною мірою і потребує додаткового дослідження і обґрунтування.

**Мета статті** полягає в обґрунтуванні поняття «система комп’ютерного моделювання» (СКМод), описі її концептуальної моделі; визначенні особливостей використання СКМод в навчальному процесі в закладів загальної середньої освіти.

## 2. МЕТОДИ ДОСЛІДЖЕННЯ

Це дослідження виконувалося в рамках науково-дослідної роботи «Система комп'ютерного моделювання пізнавальних завдань для формування компетентностей учнів з природничо-математичних предметів» (НДР №0118U003160). Під час дослідження використовувались методи аналізу педагогічної і методичної літератури, дисертаційних досліджень; узагальнення; теоретичного моделювання; системного аналізу для визначення структурних елементів моделі системи комп’ютерного моделювання; індивідуальні бесіди з учнями, вчителями та іншими категоріями педагогічних працівників; вивчення результатів навчально-пізнавальної і практичної діяльності учнів; оформлення результатів у вигляді таблиць, схем, малюнків.

## 3. РЕЗУЛЬТАТИ ДОСЛІДЖЕННЯ

Інтенсивний розвиток хмарних обчислень дав поштовх до проектування глобальних інформаційно-когнітивних систем та розробки новітніх систем, зокрема систем комп’ютерного моделювання (СКМод).

Вчені з Мерілендського університету в Коледж-Парку (UMD) і Національного інституту стандартів і технологій (NIST) США створили модель квантової системи, що складається з 53 кубітів. Квантова модель використовує кубіти для імітації складної квантової матерії. Створення моделей кубітів є ключовим завданням на шляху створення повноцінного квантового комп'ютера [10].

Лауреати Нобелівської премії з хімії М. Карплюс, М. Левітт і А. Вархель розробили комп’ютерні моделі складних хімічних систем, що стало основою для розробки комп’ютерних програм, для моделювання і передбачення складних хімічних процесів (http://www.nobelprize.org/).

Досвід провідних університетів дав підстави для висновку, що ключовим фактором перспективного розвитку науки і техніки є створення різноманітних за призначенням систем комп'ютерного моделювання, що дає можливість експериментально досліджувати, вивчати і моделювати різні природні явища і технологічні процеси.

Терміни «моделювання» в сучасному словнику іншомовних слів визначається як метод дослідження явищ і процесів, що ґрунтуються на заміні конкретного об'єкта досліджень іншим, подібними до нього – моделлю [11, с.374].

Моделювання, як метод пізнання, має багатовікову історію. Людина намагалася різними способами (словесно, графічно, за допомогою математичних формул, фізичних та технічних зразків) описати явища за якими вона спостерігала і об'єкти, що її оточували [12; 13].

Як зазначають зарубіжні дослідники перші симулятори з'явилися в XX ст. для вирішення питань підготовки пілотів для літаків.

1927-1929 рр. Е. Лінк розробив модель навігатора для переміщення по трьох осях під час управління літаком. Незабаром після цього Е. Лінк розробив імітатор з навігаційними інструментами, що дало змогу пілотам моделювати політ за приладами [14].

Потім, у 1948 р., Дослідницьким центром армії Університету Джона Хопкінса була розроблена перша імітаційна модель протиповітряної оборони. Моделювання ситуацій протиповітряної оборони здійснювалося з метою вивчення її особливостей і використання в роботі морських зенітних ракетних систем Північної Америки [15].

У 1959 р. Корпорацією Digital Equipment для монохромного міні комп'ютера, було розроблено гру – імітацію космічної війни (Space War), що могла на той час працювати в реальному часі, відрізнятися інтерактивом та графічними компонентами. У цій імітації гравцям дозволялося маневрувати кораблями, відслідковувати свої дії на круговому моніторі і використовувати ракети (Space War in Computer History Museum Lobby, www.computerhistory.org). (рис. 1)

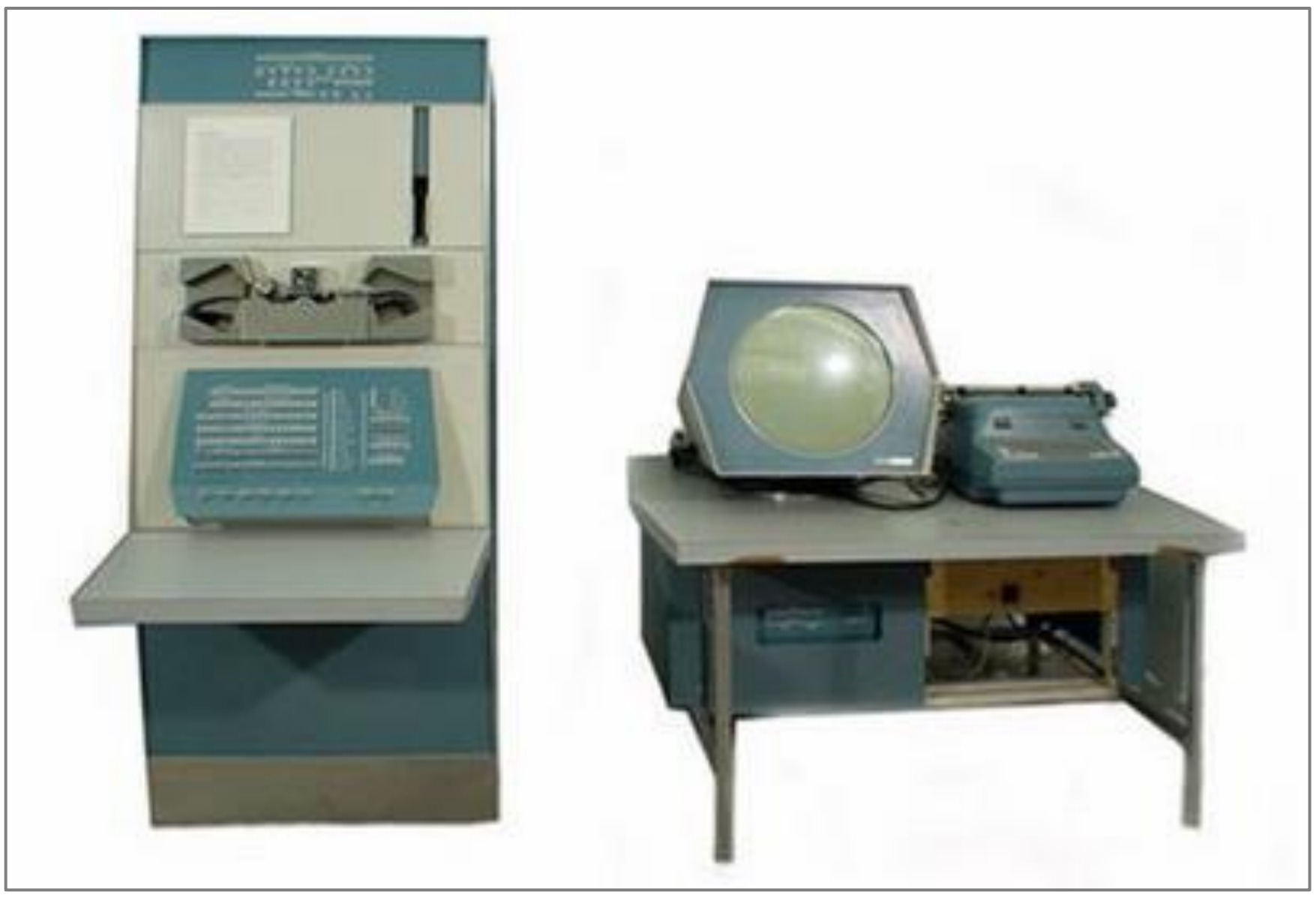

Рис. 1. Space War in Computer History Museum Lobby, 1959 р.

Нині за допомоги комп'ютерних моделей учні можуть спостерігати і експериментувати з природними явищами і процесами, які б за інших умов були б неможливі. Вчені планово розроблюють і застосовують імітаційні моделі як засоби для вивчення та поглиблення знань про природні явища в широкому діапазоні: від субатомних до планетарних.

Ці особливості роблять комп’ютерне моделювання цінним як системи освіти в цілому, так і для розуміння та прогнозування різних процесів окремо, починаючи від моделювання імунних клітин (предмет біологія), кругообігу води в природі (предмет природа), моделювання продуктів харчування (предмет хімія) до зростання населення на планеті (предмет географія).

Але, як правило, комп'ютерне моделювання пов'язують з навчанням. Його використовують для підготовки астронавтів, комерційних і військових екіпажів, фахівців з атомної енергетики, медичних працівників, пожежників, фахівців з технічного обслуговування та ін. (рис. 2-3).

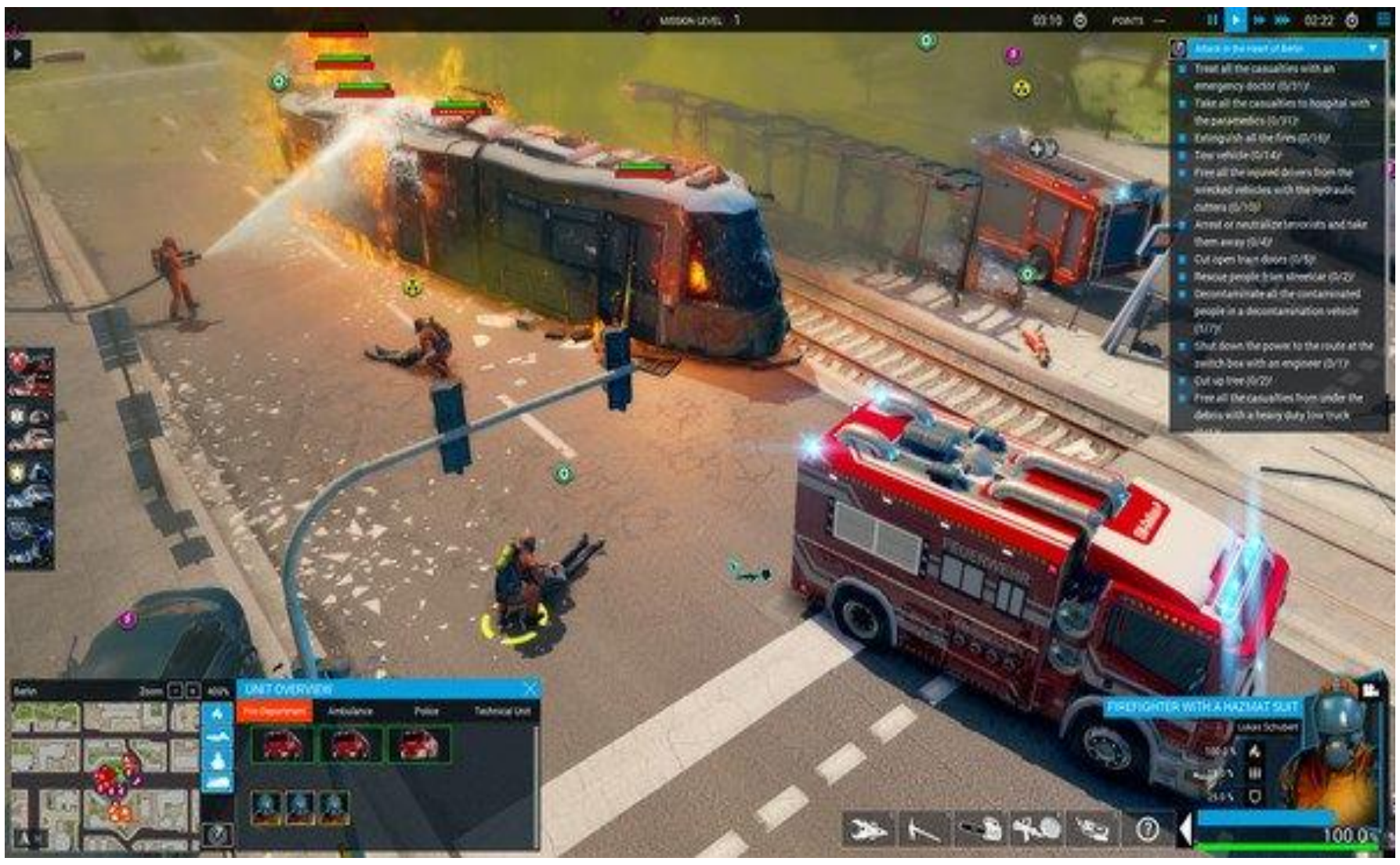

Рис. 2. Комп'ютерне моделювання ситуації «Пожежа в транспортних засобах»

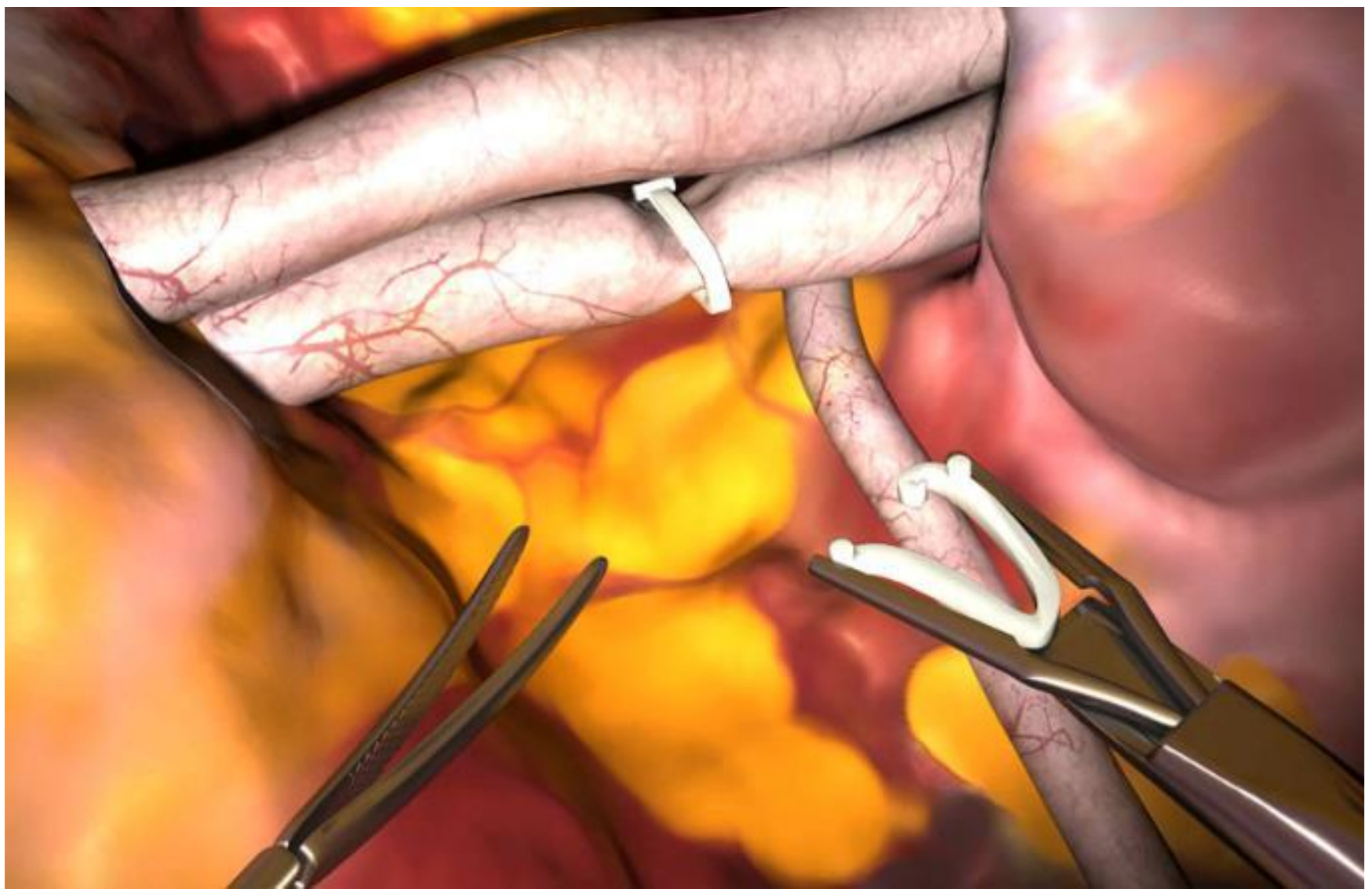

Рис. 3. Комп’ютерна імітаційна модель для підготовки хірургів

Для підготовки фахівців зазначених спеціальностей створюються такі синтетичні навчальні середовища, які допоможуть їм набути умінь, навичок та професійних компетентностей [17].

Розробка комп'ютерної моделі потребує участі фахівців високого рівня, відповідної освіти, які володіють унікальними знаннями та навичками роботи в реальних умовах [18].

Будь-яка дисципліна освітнього циклу включає базові моделі. Математики і фізики не можуть здійснювати дослідження без формул й діаграм, граматику української або зарубіжної мови неможливо освоїти без знакових конструкцій, теми з біології також важко уявити без макетів, а хімію без основних структурних моделей, зокрема комп'ютерних.

Отже, комп’ютерне моделювання - це проектування об'єкта або явища, з використанням комп’ютерної техніки та математичних, фізичних або логічних систем. Результатом иоделювання є комп’ютерні моделі, які можуть бути представлені у форматах 2D-зображення та 3D-зображення; вони можуть бути статичні або гейміфіковані.

У роботі О. М. Дахина окреслено основні етапи педагогічного моделювання, зокрема:

– вибір методології моделювання;
– вивчення властивостей предмета дослідження;
– постановка завдань;
– проектування моделі;
– визначення залежностей між основними елементами моделі;
– визначення критеріїв оцінювання моделі;
– дослідження валідності моделі у вирішенні поставлених завдань;
– застосування моделі в педагогічному експерименті;
– інтерпретація результатів моделювання [19].

У процесі проектування комп'ютерної моделі особливо важливими є:

– ідея;
– достовірність процесу;
– дизайн;
– процес реалізації;
– очікуваний результат;
– перспективи розвитку, поширення і використання [13].

Визначимо основні *функції моделювання*, а саме: дескриптивну, прогностичну і нормативну [18].

*Дескриптивна* функція полягає в тому, що за рахунок високого ступеню абстрагування на моделях просто пояснити явища і процеси. Успішні в цьому відношенні моделі стають компонентами наукових теорій і є ефективним засобом відображення змісту останніх.

*Прогностична* функція моделювання відображає можливість скласти короткостроковий або довгостроковий прогноз змін властивостей і стану змодельованих систем.

*Нормативна* функція моделювання полягає в відображенні запланованого результату. У процесі моделювання потрібно не тільки описати існуючу систему, а й побудувати її нормативний образ - бажаний з точки зору суб'єкта, який задовольняє визначеним заздалегідь критеріями.

Як зазначає Н. Данешхо (Naqib Daneshjo) комп’ютерні моделі існують двох типів [20]:

*Перший тип*. Комп’ютерні моделі, за допомоги яких можна аналізувати об’єкти або системи, перевіряти, спостерігати та уточнювати їх характеристики. Нині існує значна кількість комп’ютерних моделей природних процесів та явищ, що дозволяє здійснювати процес аналізу об’єктів простішим, пізнавальним, цікавішим та ґрунтовним.

*Другий тип*. Комп’ютерні моделі, що виникають в результаті розробки та дизайну (наприклад, модель присадибного будинку). Ця діяльність зазвичай підтримується комп'ютерними технологіями. Таким чином ми говоримо про методи автоматизації проектування, що потребує спеціального програмного забезпечення.

Процес експериментування з моделлю називатимемо *імітацією* (симуляцією) [20]. Експериментування з моделями дозволяє здійснювати глибокий аналіз та визначати характеристики об’єкта, шукати альтернативи  або розв’язки сформульованих завдань.

Враховуючи освітню складову, ми будемо розглядати такі типи *моделей*:

–*ігрові моделі* Це військовий, економічний, спортивний, рольові, відтворюючи поведінки об'єктів у різних ситуаціях, граючи їх враховуючи можливу реакцію, конкурент, союзником або ворога;

–*імітаційні моделі (симуляції)* образ об’єкта з високим ступенем відтворення його властивостей [21].

–*алгоритмічні*, як модель алгоритмічного процесу [22].

Усі зазначені комп’ютерні моделі можна об’єднати в єдину систему – систему комп’ютерного моделювання (рис. 4).

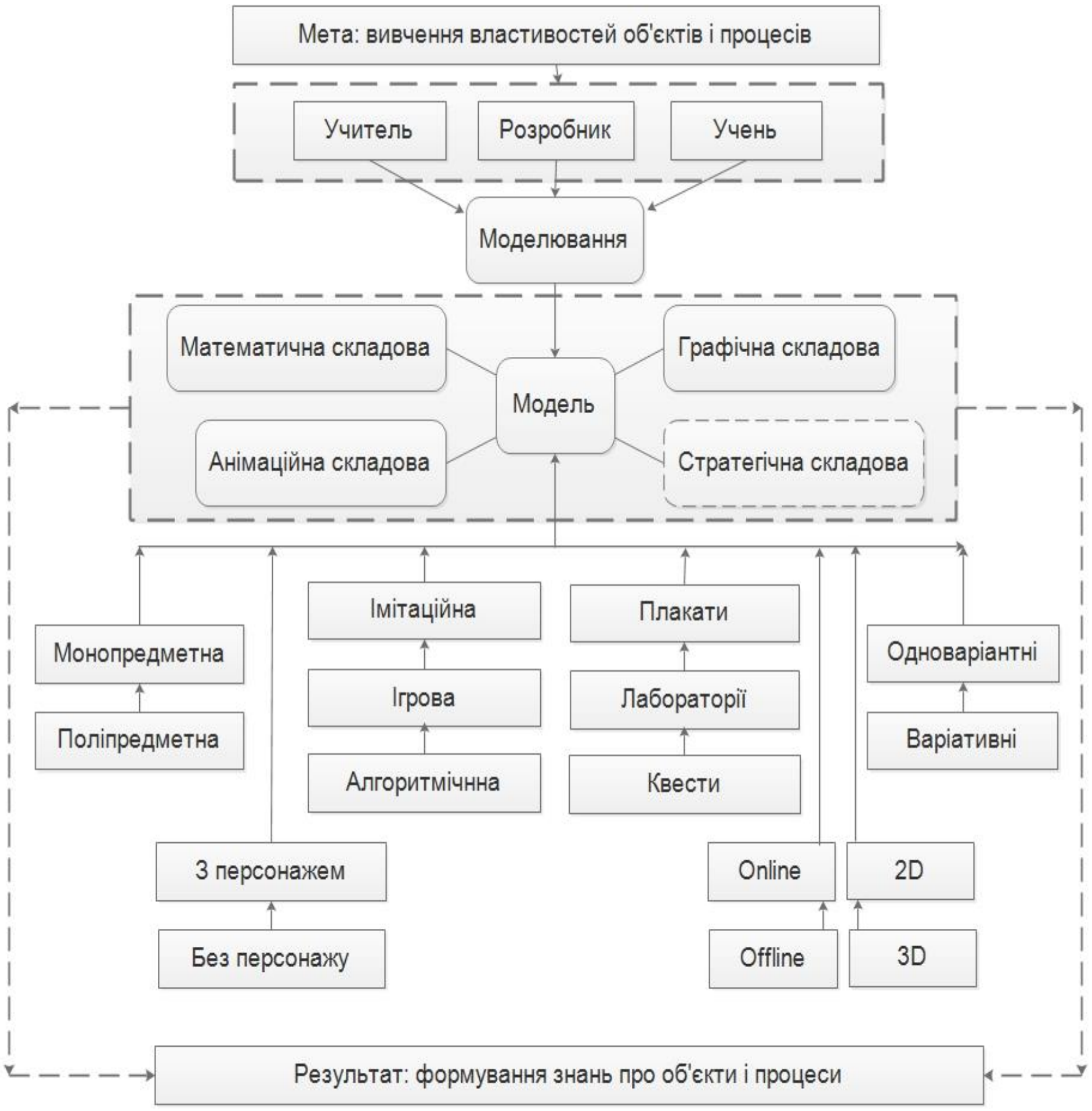

Рис. 4. Концептуальна модель СКМод

Під *системою комп'ютерного моделювання* (СКМод) будемо розуміти програмні засоби нового покоління, призначені для анімаційної візуалізації явищ і процесів, побудови стратегій дій, виконання чисельних розрахунків будь-якого рівня складності та спрямованих на унаочнення та розв'язання задач різних типів.

Розглянемо концептуальну модель СКМод і визначимо основні *види моделей, що входять до* її складу:

– монопредметні (для одного навчального предмету) і поліпредметні (одну модель можна застосувати на декількох навчальних предметів);
– імітаційні (імітація процесу або явища), ігрові (має навчальну стратегію, варіативність вибору рішень), алгоритмічні (демонстрація виконання заданого алгоритму) (рис. 5);

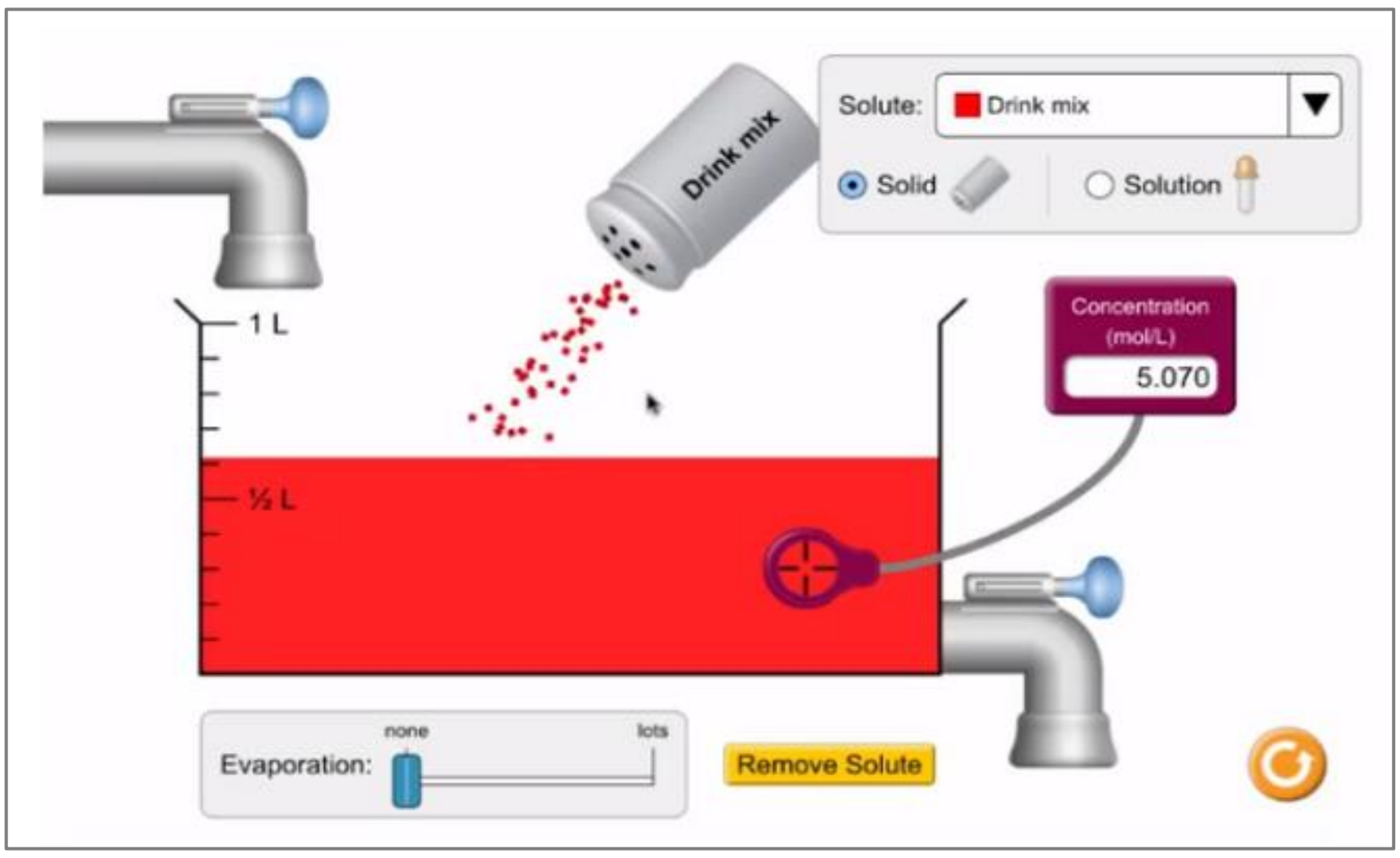

Рис. 5. Імітаційна модель, предмет хімія (СКМод Phet)

– плакати (відповідність, співставлення), лабораторії (виконання заданої послідовності дій), квести (послідовне послідовності дій з урахуванням попереднього результату);
– одноваріантні (одне завдання), варіативні(кілька завдань);
– з персонажем, без персонажу;
– онлайнові, офлайнові.

Визначимо *цілі* використання СКМод:

– створення єдиного освітнього інформаційного середовища;
– формування інформаційної культури учнів;
– формування ключових та дослідницьких компетентностей учнів з природничо-математичних предметів;
– формування індивідуальної траєкторії розвитку учня;
– підготовка учнів до самостійної навчально-пізнавальної діяльності;
– підвищення якості оволодіння знаннями.

Використання СКМод в навчанні будується на *принципах*:

– *принцип педагогічної доцільності*, що полягає в підвищенні ефективності освітнього процесу, створенні педагогічних умов як для розвитку особистості, так і для формування компетентностей учнів з природничо-математичних предметів;
– *принцип когнітивної доцільності* означає використання СКМОД для розвитку пам'яті, уваги, мислення, уяви та формування пізнавальної компетентності учнів (здатності учнів здійснювати навчально-пізнавальну діяльність);

– *принцип* дидактичної значущості, що полягає в побудові варіативних дидактичних маршрутів розвитку кожного окремого учня та формуванні дослідницької компетентності;

– *принцип* методичної ефективності, що полягає у виборі ефективність прийомів, форм і методів для вирішення освітніх завдань;

– *принцип* інтерактивної наочності для посилення ефекту візуалізації навчального контенту і створення ефекту динамічності процесів і явищ природи.

У навчальному процесі СКМод виконує такі *функції*:

– *освітню*, що полягає в забезпеченні засвоєння учнями певного обсягу знань, формуванні у них предметних компетентностей;

– *розвивальну*, що сприяє розвитку перцептивних, розумових, мовленнєвих та інших здібностей учнів;

– *виховну*, що забезпечує формування світогляду, моральних, естетичних та інших якостей особистості учня;

– *управлінську*, що полягає в програмуванні певного типу навчання, його методів, форм і засобів, способів застосування знань у різних ситуаціях;

– *дослідницьку*, що спонукає учня до самостійного вирішення проблем, навчає методів наукового пошуку;

– *візуальну*, що полягає в наочному представленні навчального матеріалу для розвитку когнітивних здібностей учнів;

– *інтерактивну*, що комбіноване представлення навчального матеріалу з використанням анімації і відео, що забезпечує активізацію навчального процесу.

Уточнимо, що складниками, що характеризують ефективність *дидактичної значущості СКМод є*: вибір оптимального змісту і структури занять; вибір найбільш раціональних методів і прийомів, а також внесення необхідних коректив в їх застосування; раціональне поєднання групових та індивідуальних форм роботи, планування витрат часу, створення сприятливих умов для самодіагностики з метою виявлення проміжних результатів навчання [23].

Розглянемо особливості використання СКМод в навчальному процесі.

*1) Організація навчального процесу*. Використанням СКМод може здійснюватися як в рамках формального (школа), так і неформального навчання (позашкільні навчальні заклади, гуртки, тематичні секції), а наявність мережі Інтернет дає можливість повсюдного доступу до СКМод в режимах онлайн і офлайн.

Використовувати СКМод можна в початковій, середній і старшій ланках загальної середньої освіти.

Наявні нині СКМод охоплюють такі навчальні предмети, як: фізика, хімія, біологія, математика.

Доцільність використання СКМод визначається вчителем і може бути включена для унаочнення нового навчального матеріалу, проведення лабораторних і практичних робіт, а також для вирішення дослідницьких, творчих і проблемних завдань.

*2) Принципи організації роботи з учнями:*

– залучати учнів до збирання даних, моделювати і спільно використовувати в навчальному процесі;

– підтримувати розвиток компетентностей в учнів;

– залучати учнів до використання СКМод, щоб допомогти їм глибше зрозуміти основи науки [23].

*3) Особливості навчання учнів.*

*Позитивні аспекти:*

– покращуються навички аналізу та прийняття рішень;

– підвищується рівень цифрової комунікації (Інтернет, мобільної, хмаро-орієнтованої та ін.);
– розвивається пильність;
– підвищується рівень навчальних досягнень з математики, письма та читання;
– навчання здійснюється через гру;
– вирішувати окремі проблеми можна навчитися у процесі використання комп'ютерних моделей, зокрема в процесі вирішення проблемних завдань;
– збільшується тривалість уваги;
– навчання технологій здійснюється природним шляхом;
– розвиваються навички стратегічного мислення і планування.

*Негативні аспекти:*

– не сприяють розвитку соціальних навичок та взаємодії з іншими;
– питання безпеки під час використання учнями мережі Інтернет залишається актуальним;
– можуть виникнути проблеми з оцінюванням навчальних досягнень учнів [23].

*4) Організація групової та індивідуальної роботи.*

Коли завдання має виконувати група учнів, завжди виникає питання розподілу обов'язків, що не сприяє позитивному клімату навчання. Тому під час використання СКМод учні мають колективно узгоджувати висновки і приймати рішення, але завдання (за бажанням) можуть виконувати індивідуально.

Так, наприклад, в СКМод EcoMUVE протягом двох тижнів учні вивчають властивості та особливості води ставка, щоб зрозуміти, чому в ньому гине риба (http://ecolearn.gse.harvard.edu/ecoMUVE/overview.php) (рис. 6).

Учні починають з того, що вивчають підводний світ ставка, вимірюють температуру води, здійснюють аналіз погодних умов, перевіряють каламутність (ясність води), періодично визначають рівні pH, працюють, щоб зрозуміти вплив основних компонентів екосистеми на стан води в ставку і пов'язаних з цими явищами причинно-наслідкові зв'язки.

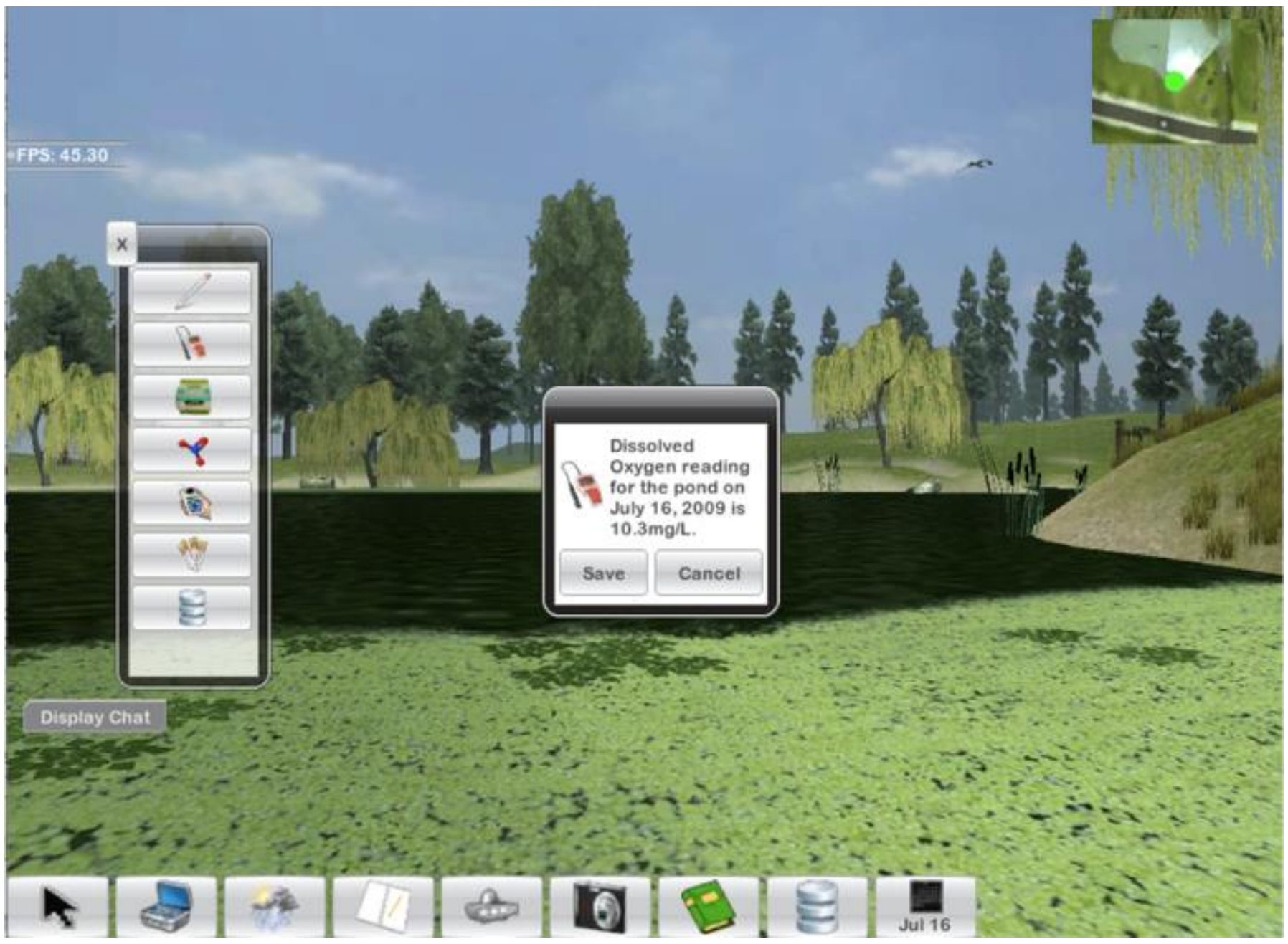

Рис. 6. Фрагмент збереження учнями даних щодо насичення озера киснем

Доречно в такій роботі спроектувати як індивідуальну роботу, так і колективні обговорення.

Підтримку активної групової роботи можна здійснювати за допомоги миттєвих повідомлень (чату), що дає додаткові можливості учням для спілкування і обговорення окремих аспектів дослідження, з'ясування проблем, пошук варіантів їх розв'язання. Дієвим буде залучення науковців до роботи учнів, які в онлайновому режимі можуть надавати консультації з тих чи інших питань дослідження.

*5) Мотивація учнів.* Важливою особливістю використання СКМод під час формального і неформального навчання є мотивація учнів. В контексті формальної освіти учень може бути як вмотивованим так і ні, а в рамках неформальної освіти учень вмотивований власними інтересами (National Research Council, 2009). Використання СКМод покращує мотивацію та інтерес учнів до вивчення предметів шкільного курсу, формує і розвиває когнітивні здатності щодо оперування абстрактними термінами і категоріями; підвищує рівень розуміння процесів живої природи і взаємозв'язків в ній [16].

*6) Формування предметних компетентностей*. Загалом СКМод дають можливість підвищити інтерес до навчання і компетентності тих учнів, які вважають предмети природничо-математичного циклу нецікавими.

Дослідження зарубіжних колег показали, що навчальний контент, що формується на наукових теоріях і підтримується СКМод може позитивно вплинути на формування предметних компетнтностей учнів [16, с.20]. Тому важливим залишається розробка системи завдань, за допомоги яких можна було б формувати індивідуальну траєкторію розвитку учнів, їх ключових та предметних компетентностей, зокрема з природничо-математичних предметів. Особливої уваги потребує розробка етапів і відбір змісту завдання для роботи учнів з СКМод (рис. 7).

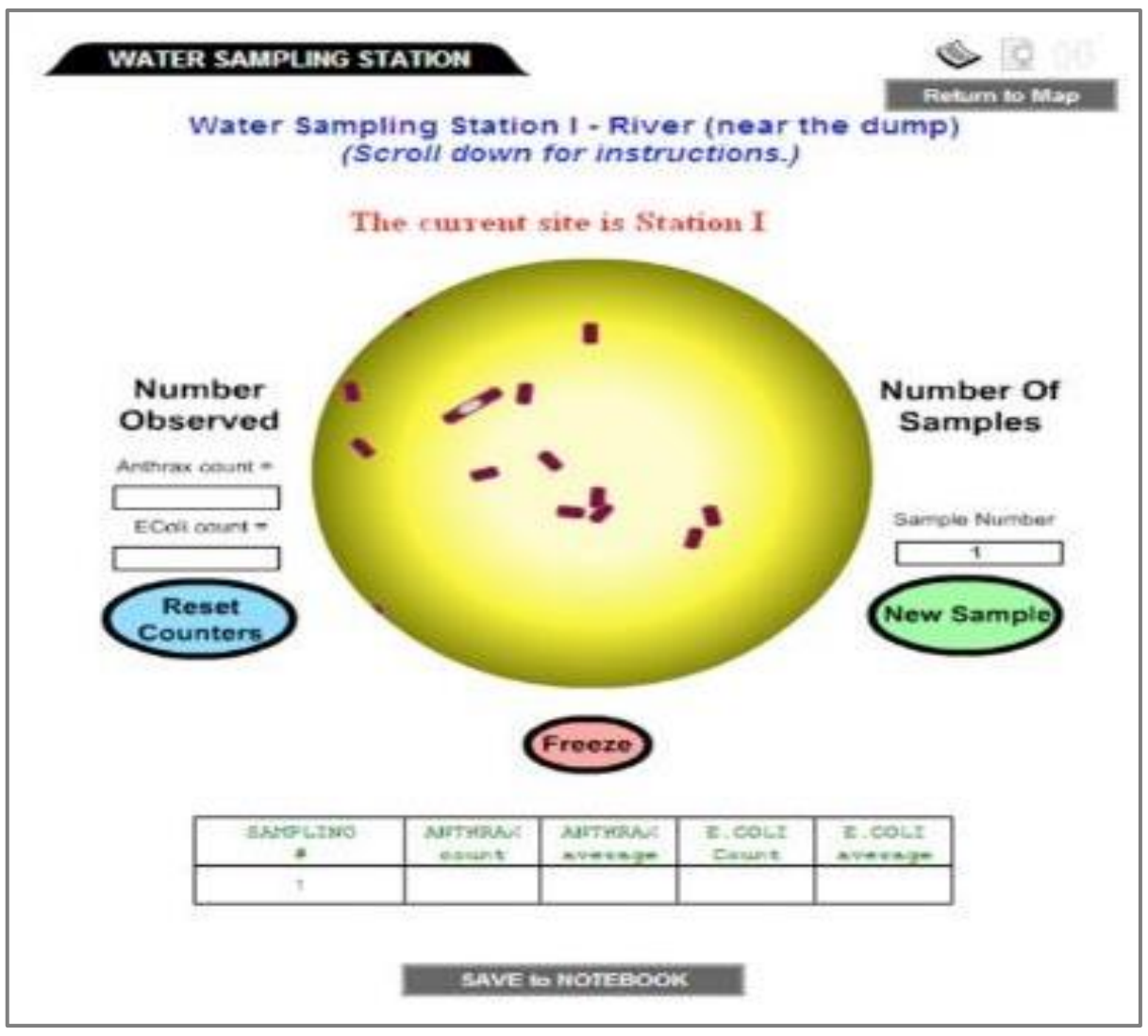

Рис. 7. Фрагмент використанням учнем віртуального мікроскопу

Забезпечення вчителів і учнів додатковими навчальними матеріалами (інструкціями, роздатковими матеріалами, робочими зошитами, планами роботи) знімає низку запитань які можуть виникнути під час роботи з СКМод.

Ще одним важливим аспектом залишається оцінювання навчальних досягнень учнів які використовують СКМОД, оскільки може статися так, що учні з високим рівнем абстрактного мислення можуть втратити інтерес до навчання.

Для таких учнів учитель має створювати додаткові навчальні ситуації, зокрема, пошук додаткових даних, визначення додаткових властивостей об'єкта тощо.

*7) Підтримка зв'язку з розробниками СКМод.*

Виробниками в шкільних класах було апробовано велику кількість різних типів комп'ютерних моделей та лише деякі з них набули широкого впровадження (рис. 8).

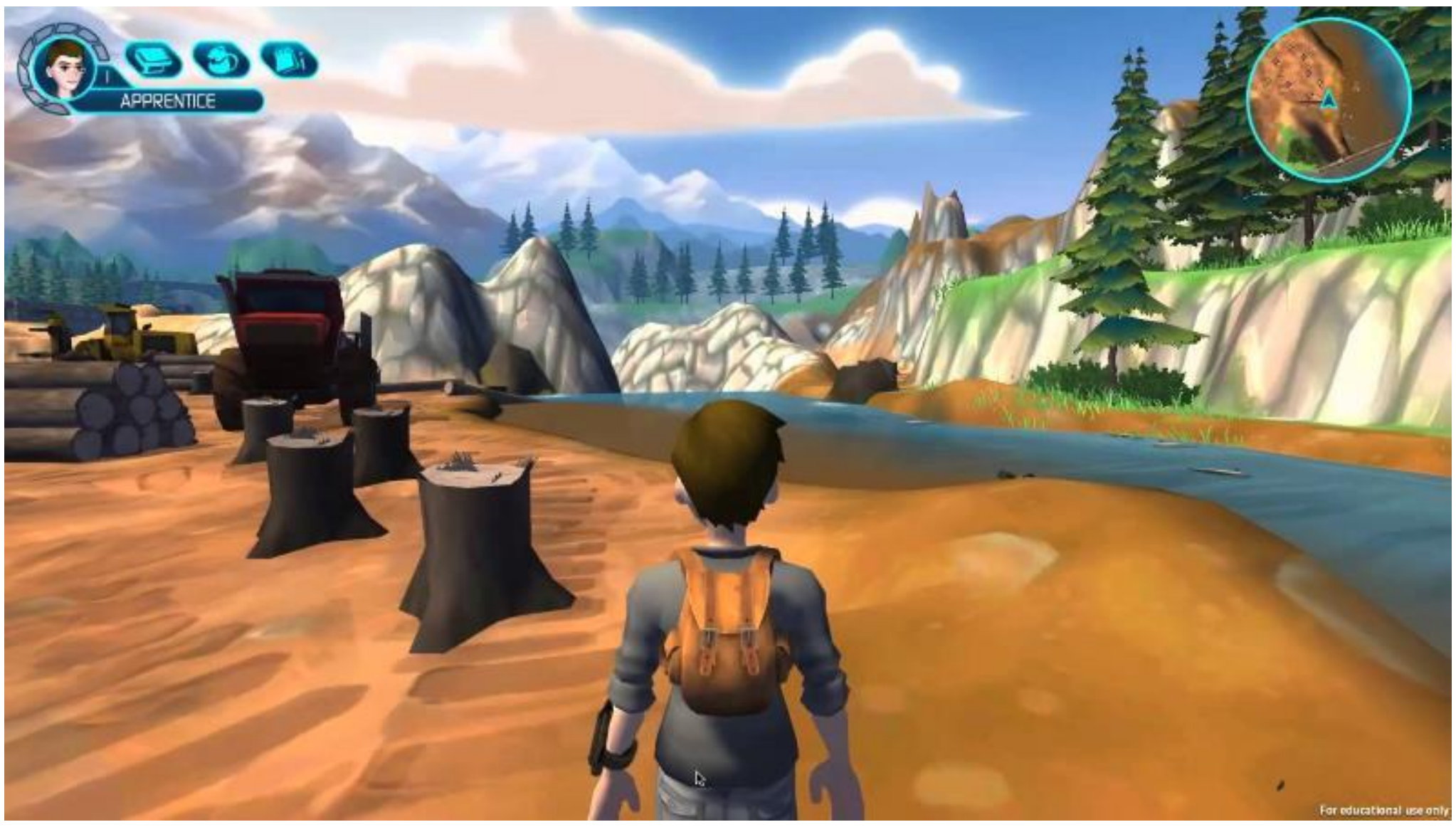

Рис. 8. Фрагмент СКМод «Mystery of Taiga River»

Наприклад СКМод «Taiga Park», яка відрізнялася цікавою ідеєю та наближеними до реального життя завданнями використовувалася під час навчання учнів початкових класів в урочний та позаурочний час протягом 2009-2013 рр. [22]. Автори СКМод моніторили використання та ефективність комп'ютерних моделей для навчання. Нині в початкових школах використовують її оновлену версію «Mystery of Taiga River».

Ця СКМод заснована на ігровому контенті, де журналістами-дослідниками стають учні, які вивчають і застосовують наукові підходи (розроблюють етапи дослідження, визначають індикатори якості води, евтрофікацію і т.д.) для вирішення проблем, пов'язаних з відновлення паркової зони з участю лісорубів, рибалок, фермерів та ін. (https://gamesandimpact.org/taiga_river/).

Ретельний відбір СКМод обумовлений реальними причинами, пов'язаними з тим, що учні знаходять можливість «обійти» алгоритми програми. Наприклад, в СКМод «River City», учитель помітив, що учні багато разів використали пастку для москітів. З'ясувалося, що на думку учнів таким чином можна зменшити популяцію москітів і заблокувати хворобу. Ця хибна гіпотеза учнів дала поштовх для удосконалення ігрової моделі.

*8) Формування інноваційного навчального середовища.*

Як зазначалося вище, використання СКМод має бути логічно вбудованим в освітній процес. Вчитель може самостійно підбирати комп'ютерні моделі і проектувати завдання для учнів. Для організації групової роботи учнів можна використати сервіси Office 365 (OneNote, Class OneNote) за допомоги яких можна сформувати інноваційне навчальне середовище в якому буде передбачено: теоретичний матеріал, опитувальники і відібрані до конкретних уроків комп'ютерні моделі [24; 25].

*9) Впровадження СКМод в навчальний процес закладів освіти.*

Використання СКМод для поліпшення якості навчання може бути реалізовано шляхом поступових еволюційних змін, а не різких змін в підходах до навчання. Важливими залишаються професійний потенціал вчителів, матеріально-технічна база закладів освіти та фінансування для широкомасштабного впровадження новітніх технологій.

Зазначимо, що СКМод мають доповнювати, унаочнювати навчальний матеріал, відповідати цілям навчання, бути доречними і забезпечувати прогнозований ефект.

Інтеграція технологій в школі має вирішальне значення для перетворення сучасної освіти. Використання інформаційно-комунікаційних технологій для показу і демонстрації не приведуть до значного поліпшення навчання природничо-математичних предметів [9, с. 65]. Замість цього можна інтегрувати СКМод в навчальний процес для забезпечення більш глибокого розуміння природних процесів та явищ, проектування дослідницької та пізнавальної діяльності, а саме: здійснювати координацію пізнавальної діяльності учнів, надавати консультації щодо реалізації складних ідей і побудови гіпотез; надавати допомогу у доборі контенту; спонукати до пошуку рішення та поглиблювати свої знання в предментній сфері, зазначає [9, с. 67].

## 4. ВИСНОВКИ ТА ПЕРСПЕКТИВИ ПОДАЛЬШИХ ДОСЛІДЖЕНЬ

СКМод – це узагальнена характеристика класу прикладних програм навчального призначення, за допомоги яких можна підсилити природничо-математичну складову освітнього процесу. Принципи функціонування і особливості використання СКМод підтверджують думку науковців, що вона є потужним засобом візуалізації природничих процесів і явищ та може суттєво допомогти учням у пізнанні навколишнього світу. А такий предмет як математика, що потребує від учня високого ступеня абстрактного мислення, може стати цікавим, зрозумілим і вивчатися як прикладна наука.

Разом з тим, зарубіжні вчені застерігають, що для отримання максимального ефекту від використання СКМод вони мають відповідати цілям і завданням навчання, а виробники мають постійно підтримувати й оновлювати СКМод до вимог часу та розвитку ІК-технологій.

До переваг СКМод можна віднести: повсюдний доступ, організацію індивідуальної, групової роботи; варіативність навчальних завдань; поліпредметність; достовірність природних процесів; високий ступінь візуалізації.

До недоліків СКМод можна віднести: відсутність методики їх використання в навчальному процесі під час здобуття початкової, базової середньої та профільної освіти; відсутність вітчизняних аналогів; кількість їх не достатня для охоплення ключових тем природничо-математичних предметів; низький рівень компетентності вчителів, що їх використовують.

Заклади освіти можуть використовувати СКМод на різних рівнях і етапах навчання та в різних контекстах, що складаються з взаємопов'язаних фізичних, соціальних, культурних і технологічних аспектів.

Використання СКМод в закладах загальної середньої освіти збільшить можливості вчителів щодо вдосконалення навчання природничо-математичних предметів. Вони зможуть індивідуалізувати процес навчання, щоб задовольнити темпи, освітні інтереси і можливості кожного конкретного учня.

Оскільки школи надають освітні послуги учням різних вікових та соціальних категорій, то широке використання СКМод потенційно може покращити доступ до високоякісних навчальних матеріалів для учнів які отримують освіту за дистанційною, заочною, мережною, екстернатною, сімейною, патронажною, індивідуальною і дуальною формами навчання та учнів з особливими потребами.

Співпраця наукових установ з виробниками може суттєво покращити ситуацію щодо розроблення стандартів та зменшення бар’єрів в розробленні й використанні СКМод закладів освіти.

Нині комп'ютерне моделювання стає новим і ефективним інструментом навчання, який забезпечує інтерактивність і інтеграцію з онлайн-ресурсами, які зазвичай були недоступні в традиційних шкільних середовищах [26].

Подальшого дослідження потребують питання використання СКМод під час навчання предметів шкільного курсу, обґрунтування аспектів формування компетентностей учнів з природничо-математичних предметів, розробка пізнавальних завдань з використанням СКМод.

## СПИСОК ВИКОРИСТАНИХ ДЖЕРЕЛ

# СИСТЕМА КОМПЬЮТЕРНОГО МОДЕЛИРОВАНИЯ И ОСОБЕННОСТИ ЕЕ ИСПОЛЬЗОВАНИЯ В УЧЕБНОМ ПРОЦЕССЕ УЧРЕЖДЕНИЙ ОБЩЕГО СРЕДНЕГО ОБРАЗОВАНИЯ

**Литвинова Светлана Григорьевна**
доктор педагогичних наук, с.н.с, заведующая отделом технологий открытой учебной среды
Институт информационных технологий и средств обучения НАПН Украины, г. Киев, Украина.
ORCID ID 0000-0002-5450-6635
*s.h.lytvynova@gmail.com*

**Аннотация.** В статье проанализирован исторический аспект формирования компьютерного моделирования как одного из перспективных направлений развития образовательного процесса. Обоснованы понятие «система компьютерного моделирования», концептуальная модель системы компьютерного моделирования (СКМод), определены ее составляющие (математическая, анимационная, графическая, стратегическая), функции, принципы и цели использования. Описаны особенности: организации работы учащихся с использованием СКМод, индивидуальной и групповой работы, формирования предметных компетентностей; рассмотрен аспект мотивации учащихся к обучению. Установлено, что учебные заведения могут использовать СКМод на разных уровнях и этапах обучения и в различных контекстах, состоящих из взаимосвязанных физических, социальных, культурных и технологических аспектов. Определено, что использование СКМод способствует индивидуализации процесса обучения, с целью удовлетворения темпов, образовательных интересов и возможностей каждого конкретного ученика. Обосновано, что применение СКМод при изучении естественно-математических предметов способствует формированию предметных компетентностей, развивает навыки анализа и принятия решений, повышает уровень цифровой коммуникации, развивает бдительность, повышает уровень знаний, увеличивает продолжительность внимания учеников. Дальнейшего исследования требует обоснование процесса формирования компетентностей учащихся по естественно-математическим предметам и проектирование познавательных задач с использованием СКМод.

# SYSTEM OF COMPUTER MODELING AND FEATURES OF THEIR USE IN THE EDUCATIONAL PROCESS OF GENERAL SECONDARY EDUCATION

**Svitlana G. Lytvynova**
Doctor of science (pedagogical sciences), senior scientific researcher,
Head of department of technology open learning environment
Institute of Information Technology and Learning Tools of the NAPS of Ukraine, Kyiv, Ukraine
ORCID ID 0000-0002-5450-6635
*s.h.lytvynova@gmail.com*

**Abstract.** The article analyzes the historical aspect of the formation of computer modeling as one of the perspective directions of educational process development. The notion of "system of computer modeling", conceptual model of system of computer modeling (SCMod), its components (mathematical, animation, graphic, strategic), functions, principles and purposes of use are grounded. The features of the organization of students work using SCMod, individual and group work, the formation of subject competencies are described; the aspect of students' motivation to learning is considered. It is established that educational institutions can use SCMod at different levels and stages of training and in different contexts, which consist of interrelated physical, social, cultural and technological aspects. It is determined that the use of SCMod in general secondary school would increase the capacity of teachers to improve the training of students in natural and mathematical subjects and contribute to the individualization of the learning process, in order to meet the pace, educational interests and capabilities of each particular student. It is substantiated that the use of SCMod in the study of natural-mathematical subjects contributes to the formation of subject competencies, develops the skills of analysis and decision-making, increases the level of digital communication, develops vigilance, raises the level of knowledge, increases the duration of attention of students. Further research requires the justification of the process of forming students' competencies in natural-mathematical subjects and designing cognitive tasks using SCMod.

## REFERENCES (TRANSLATED AND TRANSLITERATED)